\documentclass[journal=nalefd,manuscript=letter]{achemso}

\pdfoutput=1

\usepackage[active]{srcltx}
\usepackage{graphicx}
\usepackage{mathcomp}	
\usepackage{color}
\usepackage{amsmath,amssymb,amstext,graphicx,bm}
\usepackage{braket}

\author{Hannes~Watzinger}
\email{hannes.watzinger@ist.ac.at}
\affiliation[IST Austria]{Institute of Science and Technology Austria, Am Campus 1, 3400 Klosterneuburg, Austria}
\alsoaffiliation[JK University]{Johannes Kepler University, Institute of Semiconductor and Solid State Physics, Altenbergerstr. 69, 4040 Linz, Austria}

\author{Christoph~Kloeffel}
\email{c.kloeffel@unibas.ch‎}
\affiliation[University of Basel]
{University of Basel, Department of Physics, Klingelbergstr. 82, 4056 Basel, Switzerland}

\author{Lada~Vuku\v{s}i\'c}
\affiliation[IST Austria]{Institute of Science and Technology Austria, Am Campus 1, 3400 Klosterneuburg, Austria}
\alsoaffiliation[JK University]{Johannes Kepler University, Institute of Semiconductor and Solid State Physics, Altenbergerstr. 69, 4040 Linz, Austria}

\author{Marta D.~Rossell}
\affiliation[Empa]
{Electron Microscopy Center, Empa, Swiss Federal Laboratories for Materials Science and Technology, \"Uberlandstrasse 129, 8600 D\"ubendorf, Switzerland}
\alsoaffiliation[IBM Zurich] {IBM Research Z\"urich, CH-8803 R\"uschlikon, Switzerland}

\author{Violetta~Sessi}
\affiliation
{Technical University Dresden, Chair for Nanoelectronic Materials, 01062 Dresden, Germany}

\author{Josip~Kuku\v{c}ka}
\affiliation[IST Austria]{Institute of Science and Technology Austria, Am Campus 1, 3400 Klosterneuburg, Austria}
\alsoaffiliation[JK University]{Johannes Kepler University, Institute of Semiconductor and Solid State Physics, Altenbergerstr. 69, 4040 Linz, Austria}

\author{Raimund~Kirchschlager}
\affiliation[IST Austria]{Institute of Science and Technology Austria, Am Campus 1, 3400 Klosterneuburg, Austria}
\alsoaffiliation[JK University]{Johannes Kepler University, Institute of Semiconductor and Solid State Physics, Altenbergerstr. 69, 4040 Linz, Austria}

\author{Elisabeth~Lausecker}

\affiliation[IST Austria]{Institute of Science and Technology Austria, Am Campus 1, 3400 Klosterneuburg, Austria}
\alsoaffiliation[JK University]{Johannes Kepler University, Institute of Semiconductor and Solid State Physics, Altenbergerstr. 69, 4040 Linz, Austria}

\author{Alisha~Truhlar}
\affiliation[IST Austria]{Institute of Science and Technology Austria, Am Campus 1, 3400 Klosterneuburg, Austria}
\alsoaffiliation[JK University]{Johannes Kepler University, Institute of Semiconductor and Solid State Physics, Altenbergerstr. 69, 4040 Linz, Austria}

\author{Martin~Glaser}
\affiliation[JK University]
{Johannes Kepler University, Institute of Semiconductor and Solid State Physics, Altenbergerstr. 69, 4040 Linz, Austria}

\author{Armando~Rastelli}
\affiliation[JK University]
{Johannes Kepler University, Institute of Semiconductor and Solid State Physics, Altenbergerstr. 69, 4040 Linz, Austria}

\author{Andreas~Fuhrer}
\affiliation[IBM Zurich]
{IBM Research Z\"urich, CH-8803 R\"uschlikon, Switzerland}

\author{Daniel~Loss}
\affiliation[University of Basel]
{University of Basel, Department of Physics, Klingelbergstr. 82, 4056 Basel, Switzerland}

\author{Georgios~Katsaros}
\affiliation[IST Austria]{Institute of Science and Technology Austria, Am Campus 1, 3400 Klosterneuburg, Austria}
\alsoaffiliation[JK University]{Johannes Kepler University, Institute of Semiconductor and Solid State Physics, Altenbergerstr. 69, 4040 Linz, Austria}

\title{Heavy hole states in Germanium hut wires}

\begin{document}


\begin{abstract}

Hole spins have gained considerable interest in the past few years due to their potential for fast electrically controlled qubits.
Here, we study holes confined in Ge hut wires, a so far unexplored type of nanostructure. Low temperature magnetotransport measurements reveal a large anisotropy between the in-plane and out-of-plane g-factors of up to 18. Numerical simulations verify that this large anisotropy originates from a confined wave function which is of heavy hole character. 
A light hole admixture of less than 1\% is estimated for the states of lowest energy, leading to a surprisingly large reduction of the out-of-plane g-factors. 
However, this tiny light hole contribution does not influence the spin lifetimes, which are expected to be very long, even in non isotopically purified samples.

\end{abstract}

\begin{figure*}
	\center
	\includegraphics[scale = 1]{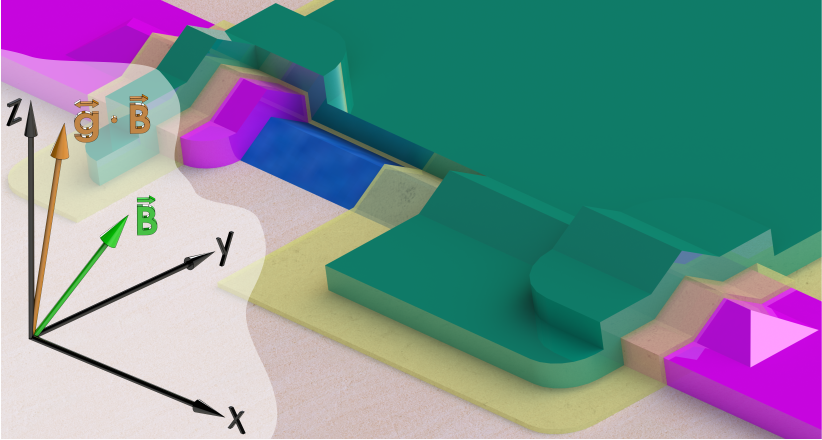}
\end{figure*}

\newpage

The interest in group IV materials for spin qubits has been continuously increasing over the past few years after the demonstration of long electron spin lifetimes and dephasing times \cite{Morello2010, Maune2012, Buech2013, Simmons2011, Zwanenburg2013}. Silicon (Si) has not only the advantage of being the most important element in semiconductor industry, it can be also isotopically purified eliminating the problem of decoherence from hyperfine interactions. Indeed, the use of such isotopically purified samples allowed the observation of electron spin coherence times of almost a second \cite{Muhonen2014}. One limitation of Si is the difficulty to perform fast gate operations while maintaining the good coherence. One way around this problem is to use the spin-orbit interaction of holes \cite{Li2015} and tune the spin with electric fields. First steps in this direction have been recently reported\cite{Maurand2016}. Holes in Germanium (Ge) have an even stronger spin-orbit coupling \cite{Kloeffel2011,Hao2010, Higginbotham2014PRL}. This fact together with the rather weak hyperfine interaction, already in non purified materials, make Ge quantum dots (QDs) a promising platform for the realization of high fidelity spin qubits \cite{Kloeffel:PRB2013}.

In 2002 the first Ge/Si core shell nanowires (NWs) were grown by chemical vapor deposition\cite{Lauhon2002} and soon after, QDs were investigated in such structures \cite{Roddaro2008, Lu2005, Hu2007}. The cylindrical geometry of the NWs, however, leads to a mixture of heavy holes (HH) and light holes (LH) \cite{Kloeffel2011, sercel:prb90, csontos:prb09}.
As a consequence, the hyperfine interaction is not of Ising type, which thus reduces the spin coherence times \cite{Fischer2008}. Still, spin relaxation times of about $600\;\mu$s \cite{Hu2012} and dephasing times of about 200\;ns\cite{Higginbotham2014} were reported. A way of creating Ge QDs with non-cylindrical symmetry is by means of the so called Stranski-Krastanow (SK) growth mode\cite{Stangl2004}. In 2010, the first single hole transistors based on such SK Ge dome-like nanostructures were realized \cite{Katsaros2010}. Electrically tunable g-factors were reported \cite{Ares2013PRL} and Rabi frequencies as high as 100\,MHz were predicted \cite{Ares2013APL}.  However, due to their very small size it is difficult to create double QD structures, typically used in spin manipulation experiments \cite{Koppens2006}.  
A solution to this problem can come from a second type of SK Ge nanostructures, the hut clusters, which were observed for the first time in 1990 \cite{Mo1990}. Zhang et al.\cite{Zhang2012} showed in 2012 that under appropriate conditions the hut clusters can elongate into Ge hut wires (HWs) with lengths exceeding one micrometer. Two years later, also the growth of SiGe HWs was demonstrated\cite{Watzinger2014}. HWs have a triangular cross section with a height of about 2\,nm above the wetting layer (WL) and are fully strained. These structural properties should lead to a very large HH-LH splitting minimizing the mixing and as a consequence the non Ising type coupling to the nuclear spins. Despite this interesting perspective, not much is known about the electronic properties. 

Here, we study three-terminal devices fabricated from Ge HWs. Scanning transmission electron microscopy (STEM) images verify that during their formation via annealing no defects are induced. From magnetotransport measurements a strong in-plane versus out-of-plane g-factor anisotropy can be observed and numerical simulations reveal that the low-energy states in the HWs are of HH type. The calculated results are consistent with the experimental data and confirm that confined holes in Ge are promising candidates for spin qubits.

\begin{figure*}
	\center
	\includegraphics[scale = 1]{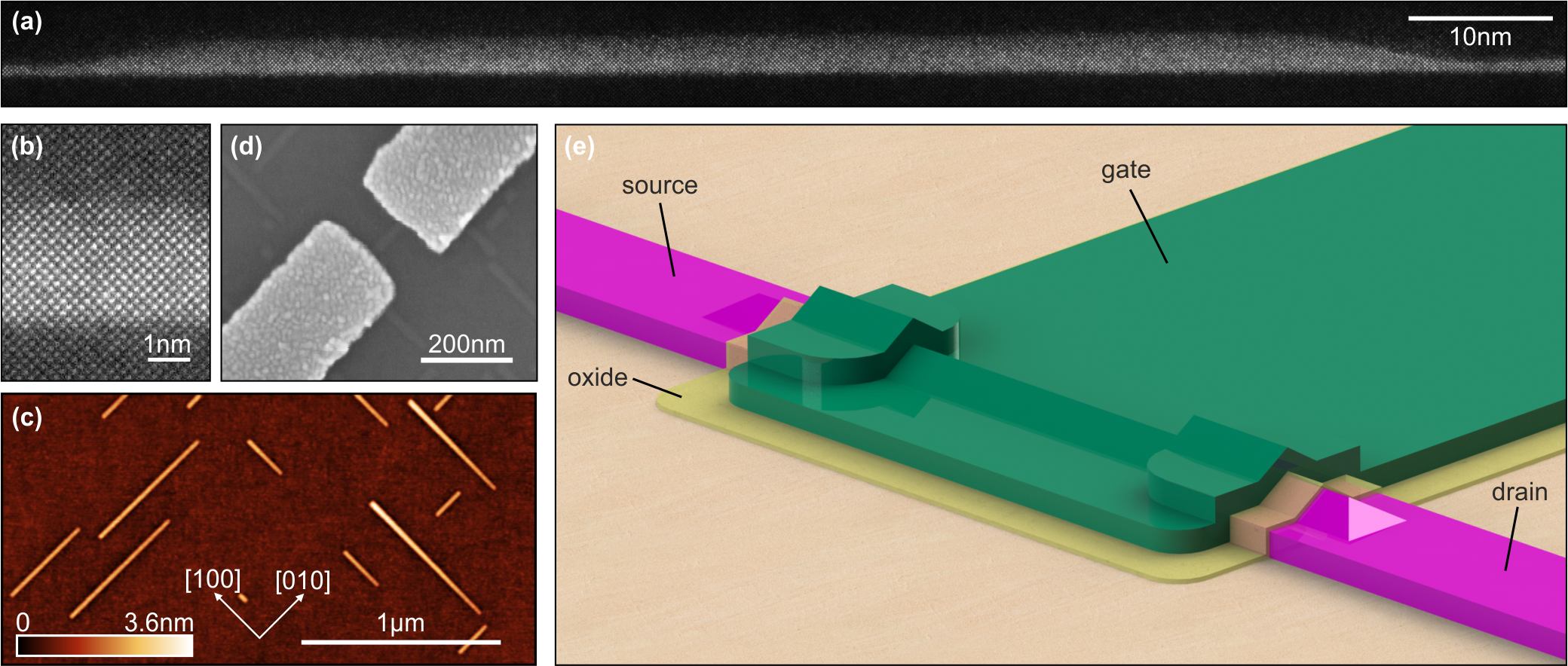}
	\caption{ (a) Scanning transmission electron microscope image along a HW embedded in epitaxial silicon. (b) Wire cross section at higher resolution showing the defect-free growth of the wires. 
(c) Atomic force microscopy image of uncapped Ge HWs. (d) Scanning electron micrograph of a HW contacted by Pd source and drain electrodes. (e) Schematic representation of a processed three-terminal device studied in this work. The Ge HW which is grown on a Si substrate and its source and drain electrodes are covered by a thin hafnium oxide layer. The top gate covers the HW and partly the source and drain contacts. } 
	\label{fig::ImageI}
\end{figure*}

The Ge HWs used in this study were grown by means of molecular beam epitaxy on 4 inch low miscut Si(001) wafers as described in Ref. \citenum{Watzinger2014}.  
6.6\,{\AA} of Ge were deposited on a Si buffer layer, leading to the formation of hut clusters. 
After a subsequent annealing process of roughly three hours, in-plane Ge HWs with lengths of up to 1 micrometer were achieved. In the last step of the growth process, the wires were covered with a 5\,nm thick Si cap to prevent the oxidation of Ge.
Figure \ref{fig::ImageI} (a) shows a STEM image taken with an annular dark-field detector. The Ge HW and the WL (bright) are surrounded by the Si substrate below and the Si cap on top (dark). The STEM lamella containing the HW was prepared along the [100] direction by focused ion beam milling and thinned to a final thickness of about 60\;nm. The TEM images show no signs of dislocations or defects, indicating perfect heteroepitaxy [see also Fig. \ref{fig::ImageI} (b)]. The height of the encapsulated wires is about 20 monolayers (2.8\;nm), including the WL. Besides having well-defined triangular cross sections, the HWs are oriented solely along the [100] and the [010] direction as can be seen in the atomic force micrograph of uncapped Ge HWs in Figure \ref{fig::ImageI} (c). 
\newline
For the fabrication of three-terminal devices, metal electrodes were defined by electron beam lithography. 
After a short oxide removal step with buffered hydrofluoric acid, 30\,nm thick palladium (Pd) contacts were evaporated. The gap between source and drain electrodes 
ranges from 70 to 100\,nm and is illustrated in Figure \ref{fig::ImageI} (d). 
The sample was then covered by a 10-nm-thick hafnium oxide insulating layer. As a last step, top gates consisting of Ti/Pd 3/20\,nm were fabricated.
A schematic representation of a processed HW device is depicted in Figure \ref{fig::ImageI} (e).

\begin{figure}
	\center
	\includegraphics[scale = 1]{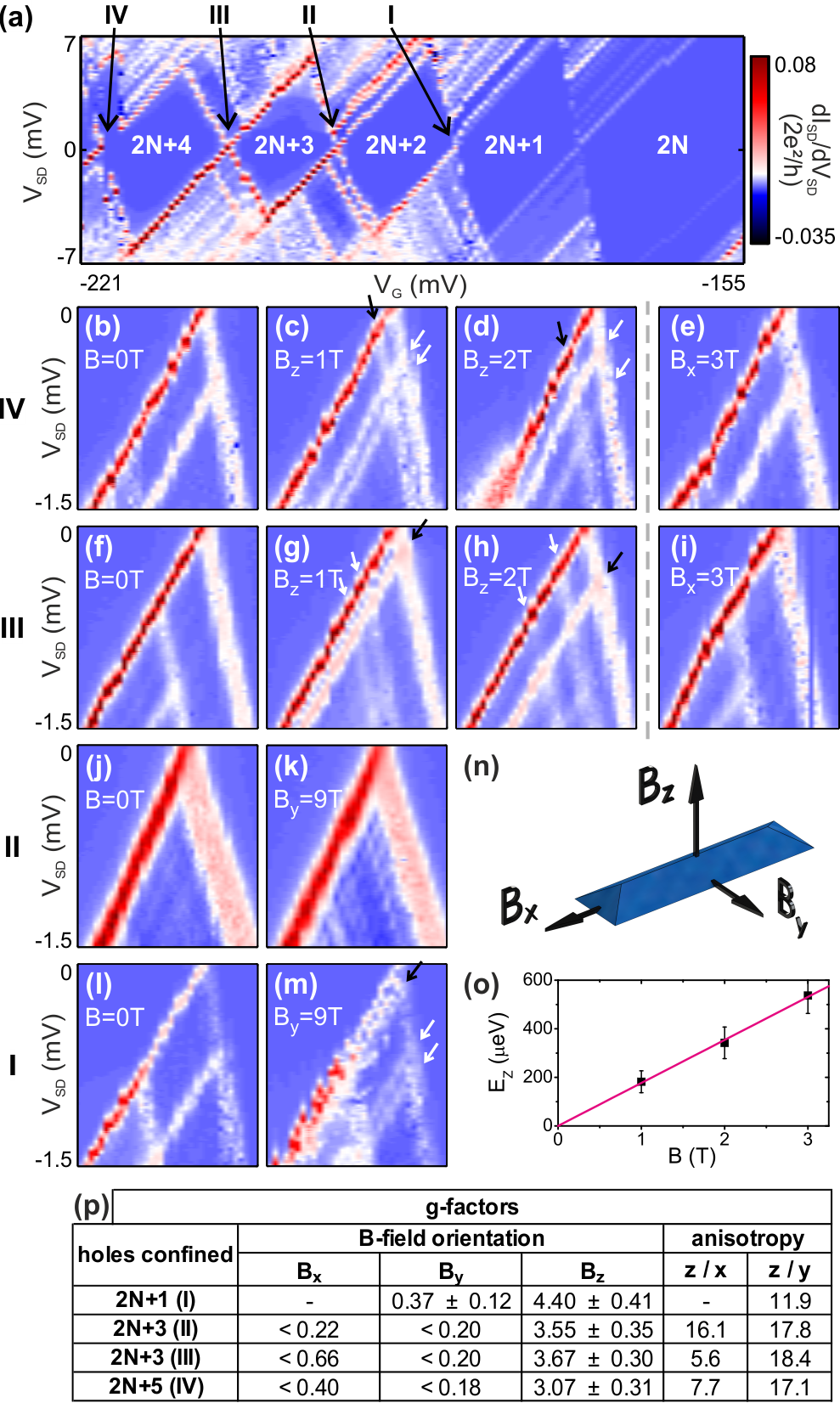}
	\caption{(a) Stability diagram of a HW device taken at  $\approx$\,250\,mK and zero magnetic field. The number of confined holes is indicated in white and the relevant crossings are labeled with roman numerals. (b-e) Differential conductance measurements versus $V_\textrm{G}$ (x-axis) and $V_\textrm{SD}$ (y-axis) for crossing IV and $B_\textrm{z} =$ 0, 1 and 2\,T, and $B_\textrm{x} =$ 3\,T, respectively. Similarly, (f-h) shows the differential conductance of the lower half of crossing III versus $V_\textrm{G}$ and $V_\textrm{SD}$ for $B_\textrm{z} =$ 0, 1 and 2\,T and (i) for $B_\textrm{x} =$ 3\,T. Measurements of crossing II are shown in (j) for 0\,T and in (k) for $B_\textrm{y} =$ 9\,T. Likewise, (l) and (m) show the lower part of crossing I at 0\,T and $B_\textrm{y} =$ 9\,T, respectively. For all measurements shown in (b-m) the gate range is roughly 6\,mV. In (n) the used nomenclature for the magnetic field orientations is illustrated. (o) Dependence of the Zeeman energy $E_\textrm{Z}$ of the GS in crossing IV versus $B_\textrm{z}$. The g-factors are extracted from the linear fit (red line).  The measured g-factors for the three different magnetic field orientations as well as the resulting anisotropies z/x and z/y are listed in (p) for crossings I to IV.} 
	\label{fig::ImageII}
\end{figure}

The devices were cooled down in a liquid He-3 refrigerator with a base temperature of about 250\,mK equipped with a vector magnet. The sample characterization was performed using low noise electronics and standard lock-in techniques.
\newline
In the following, the results of two similar devices are presented that only differ slightly in the gap size between source and drain; the two devices have channel lengths of 95\,nm and 70\,nm, respectively. A stability diagram of the first device is shown in Figure \ref{fig::ImageII} (a). Closing Coulomb diamonds prove a single QD to be formed in the HW.
Typical charging energies lie between 5 and 10\,meV and excited states (ES) can be clearly observed. The corresponding level spacing between the ground states (GS) and the first ES is up to 1\,meV.
Since at more positive gate voltages the current signal becomes too small to be measured, we cannot define the absolute number of holes confined in the QD. In order to get additional information, the device was cooled down and measured at 4\,K by RF reflectometry \cite{Reilly2007}. The reflectometry signal did not reveal the existence of additional holes beyond the regime where the current signal vanished. Thus, we estimate that in the discussed crossings the number of holes is about 20, i.e. the QD states form most likely from the first subband.
\newline
For holes the band structure is more complex than for electrons. At the $\Gamma$-point, the HH and LH bands are degenerate. This degeneracy can be lifted by strain and confinement \cite{Davies}.
The HH states in compressively strained two-dimensional hole gases lie lower in energy than the LH states, making them energetically favorable \cite{Haendel2006}. However, further carrier confinement can induce a strong mixture of HH and LH states\cite{Nenashev2003}. 
\newline
In order to investigate the nature of the HW hole states, their g-factors were determined via magnetotransport measurements. In the presence of an external magnetic field B the doubly degenerate QD energy levels split. For more than 15 diamond crossings the Zeeman splitting was measured for the three orientations illustrated in Figure \ref{fig::ImageII} (n). 
In Figure \ref{fig::ImageII} (a-m), measurements of four representative crossings showing the differential conductance ($\textrm{d}I_\textrm{SD}/\textrm{d}V_\textrm{SD}$) versus gate ($V_\textrm{G}$) and source-drain voltage ($V_\textrm{SD}$) at various magnetic fields are presented. The signature of a singly occupied doubly degenerate level is the appearance of an additional line ending at both sides of the diamond once a magnetic field is applied. These extra lines are indicated by black arrows in Figure \ref{fig::ImageII} (c) and (d) for crossing IV, in (g) and (h) for crossing III and in (m) for crossing I. They allow us to identify the diamonds between crossing II and III and on the right side of crossing I as diamonds with an odd number of confined holes. 
\newline
In addition, from the position of these extra lines the Zeeman energy $E_\textrm{Z}=g \mu_BB$ can be extracted with $\mu_B$ the Bohr magneton and $g$ standing for the absolute value of the g-factor. By plotting the Zeeman energies versus the magnetic field and by applying a linear fit to the data, the hole Land\'e  g-factor can be determined [see Figure \ref{fig::ImageII} (o)]. For crossing IV and an out-of-plane magnetic field we determine $g_{\perp}=3.07\pm0.31$. The same type of measurements result in a slightly higher value of the $g_{\perp}$- factor for the diamonds with a smaller amount of holes.  
Compared to the out-of-plane magnetic field, the in-plane directions have an almost negligible effect on the hole state splitting as shown in Figure \ref{fig::ImageII} (e) for crossing IV and in (i) for crossing III, both at $B_\textrm{x} =$ 3\,T. Due to the thermal broadening, the split lines can be barely resolved. Therefore, an upper limit of the g-factor is given for these cases. The lower parts of crossings II and I at $B_\textrm{y} =$ 9\,T are shown in Figure \ref{fig::ImageII} (k) and (m), respectively, where only the latter shows an observable splitting.
The small g-factors for both in-plane magnetic fields lead to large g-factor anisotropies z/x and z/y ranging from 5 to about 20 as shown in the table in Figure \ref{fig::ImageII} (p). A similar anisotropy was observed in crossing IV (III) for the triplet splitting indicated by white arrows in Figure \ref{fig::ImageII} (c) [(g)] and (d) [(h)] resulting in $g_{\perp}=2.61\pm0.56$; the corresponding in-plane splitting is too small to be resolved at 250\,mK. Comparing the measured g-factors with those reported for Ge dome-like QDs \cite{Katsaros2010, Ares2013APL}, it is observed that HWs have larger $g_{\perp}$ and much larger anisotropies, which are both characteristics of HH states \cite{Haendel2006}.

\begin{figure}
	\center
	\includegraphics[scale = 1]{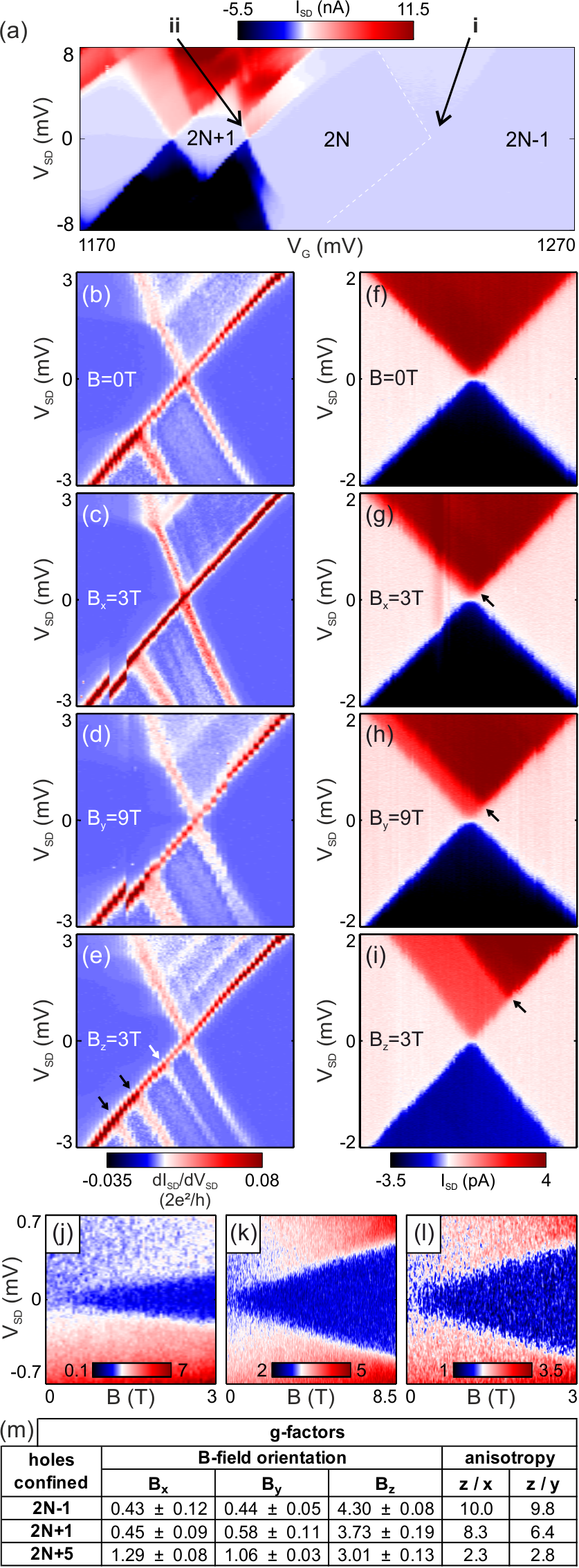}
	\caption{ (a) Stability diagram of the second device with a focus on the crossings denoted as i and ii. The magnetic field dependence is shown in (b-e) for crossing ii and in (f-i) for crossing i. For crossing ii also the splitting of the excited state can be observed as indicated in (e) by black arrows. The corresponding g-factors were extracted to $g_{\perp}=3.79\pm 0.45$ and $g_{\parallel}<1.30$ along x and $g_{\parallel}<0.68$ along y. (j-l) show differential conductance plots of inelastic cotunneling measurements for the 2N+5 hole state versus $V_\textrm{SD}$ and the magnetic field for $B_\textrm{x}$, $B_\textrm{y}$ and $B_\textrm{z}$ from left to right. The color scale insets indicate the differential conductance in units of  2$e^2$/h $\cdot 10^{-4}$. 
In (m), the determined g-factor values and the corresponding anisotropy factors for the ground state of the discussed crossings are listed. The g-factors were determined from direct tunneling except the values for 2N+1 holes at $B_\textrm{x} =$ 3\,T and for 2N+5 holes which were obtained from inelastic cotunneling measurements.} 
	\label{fig::ImageIII}
\end{figure}

In order to validate whether our findings are general characteristics of HW devices, also a second device was measured. 
Figure \ref{fig::ImageIII} (a) shows the corresponding overview stability diagram with a focus on crossing i and ii. 
Due to reasons of visibility, the corresponding magnetic field spectroscopy measurements are partly shown in current representation. In Figure \ref{fig::ImageIII} (b-e) and (f-i) the dependence on the three different B-field orientations is illustrated for crossing ii and i, respectively. Inelastic cotunneling measurements for 2N+5 holes are shown in Figure \ref{fig::ImageIII} (j), (k) and (l) in dependence of $B_\textrm{x}$, $B_\textrm{y}$ and $B_\textrm{z}$, respectively. 
The obtained g-factors are listed in the table in Figure \ref{fig::ImageIII} (m) with the highest out-of-plane g-factor being 4.3, similar to the first device. For the in-plane g-factors, slightly increased values can be observed.
However, the g-factors in out-of-plane direction are still 10 times larger than for the in-plane orientation.

From the listed g-factor values, two interesting observations can be made. a) As for the first device, the $g_{\perp}$-factor is decreasing for a higher number of holes and b) the $g_{\parallel}$-factors have clearly increased for a larger number of holes. As a consequence, a decrease of the anisotropies to less than 3 was observed for the 2N+5 hole state, indicating a more LH state.

\begin{figure}
\center
\includegraphics[scale = 1]{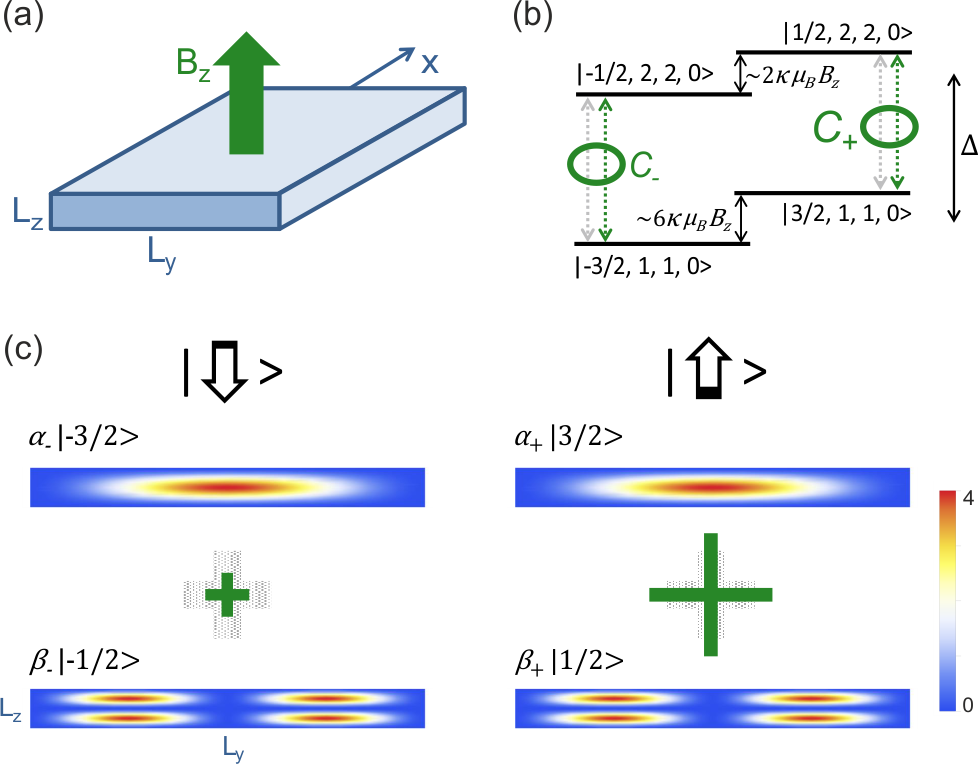}
	\caption{ (a) Sketch of the HW model in the theoretical analysis. The cross section is approximated by a rectangle of width $L_y$ and small thickness $L_z$. The green arrow represents an out-of-plane magnetic field $B_z$. (b) Effective four-level system used to derive the dominant correction $g_C$ [Eq.~(\ref{eq:gC})] in the out-of-plane g-factor $g_\perp \simeq 6 \kappa + 27 q /2 + g_C$. The LH states $\ket{\pm 1/2, 2, 2, 0}$ and the HH states $\ket{\pm 3/2, 1, 1, 0}$ [see Eqs.~(\ref{eq:notationBasisStates}) and (\ref{eq:functionOrbitalPart}) for details] differ by an energy of order $\Delta$. In the presence of $B_z$, existing couplings of equal strength (gray arrows) are reduced and enhanced, respectively, which results in $\left|C_{-}\right| < \left|C_{+}\right|$ for $B_z > 0$ as sketched in the diagram. (c) The Zeeman-split eigenstates of lowest energy after diagonalization of the system in (b). The ground state $\alpha_{-} \ket{-3/2, 1, 1, 0} + \beta_{-} \ket{-1/2, 2, 2, 0}$ (left, pseudo-spin down) consists of a HH state with spin $\ket{-3/2}$ whose probability density has a peak at the center of the HW cross section and a LH state with spin $\ket{-1/2}$ and four peaks near the corners (analogous for the excited state shown on the right, pseudo-spin up). The plots for the probability densities are dimensionless and correspond to $L_z L_y \left|\varphi_{1, 1, 0}\right|^2$ and $L_z L_y \left|\varphi_{2, 2, 0}\right|^2$, respectively [Eq.~(\ref{eq:functionOrbitalPart})]. We find $\left|\alpha_{\pm}\right|^2 > 0.99$ for typical parameters, so the LH admixtures are very small. However, due to $\left|C_{-}\right| < \left|C_{+}\right|$ caused by $B_z$, the LH admixtures $\left|\beta_{-}\right|^2 < \left|\beta_{+}\right|^2$ differ slightly, as illustrated by the different plus signs (green) and the different LH contributions (black, not to scale) in the arrows for the pseudo-spin. This difference is associated with a substantial reduction of $g_\perp$, see $g_C$. The gray plus signs of equal size in the background refer to the initial couplings which are reduced or enhanced, respectively, in the presence of $B_z$. } 
	\label{fig::ImageTheory}
\end{figure}

In order to get a better understanding of the measured g-factor values and their anisotropies we consider a simple model for hole states in HWs. Taking into account the HH and LH bands of Ge and assuming that the HWs are free of shear strain, our model Hamiltonian in the presence of a magnetic field is
\begin{eqnarray}
H &=& \frac{\hbar^2}{2m}\left[ 
\left(\gamma_1 + \frac{5 \gamma_2}{2}\right)k^2 
- 2 \gamma_2 \sum_\nu k_\nu^2 J_\nu^2
- 4 \gamma_3 \left( \{k_x, k_y\}\{J_x, J_y\} \mbox{ + c.p.} \right)
\right] \nonumber \\
& & + 2 \mu_B \bm{B} \cdot \left( \kappa \bm{J} + q \bm{\mathcal{J}} \right) 
+ b \sum_\nu \epsilon_{\nu\nu} J_\nu^2 + V(y,z). 
\label{eq:ModelHamiltonianStart}
\end{eqnarray}
It comprises the Luttinger-Kohn Hamiltonian \cite{luttinger:pr56}, the Bir-Pikus Hamiltonian \cite{Comment:Bir-Pikus, birpikus:book}, and the confinement in the transverse directions $V(y,z)$, for which we take a rectangular hard-wall potential of width $L_y$ and height $L_z$ for simplicity, i.e., $V(y,z) = 0$ if both $|y| < L_y/2$ and $|z| < L_z/2$ and $V(y,z) = \infty$ otherwise. We note that $-H$ refers to the valence band electrons, and a global minus was applied for our description of holes (which are removed valence band electrons). In Eq.~(\ref{eq:ModelHamiltonianStart}), $\{A, B\} = (AB + BA)/2$, ``c.p.'' are cyclic permutations, $m$ is the bare electron mass, and $J_\nu$ are dimensionless spin-3/2 operators. The subscript $\nu$ stands for the three axes $x$, $y$, $z$, which are oriented along the length, width, and height, respectively, of the HW [see Figures \ref{fig::ImageII}~(n) and \ref{fig::ImageTheory}~(a)] and coincide with the main crystallographic axes. With the listed vector components referring to the unit vectors along these three directions, the magnetic field is $\bm{B} = (B_x, B_y, B_z)$ and furthermore $\bm{J} = (J_x, J_y, J_z)$, $\bm{\mathcal{J}} = (J_x^3, J_y^3, J_z^3)$. The operators $k_\nu$ are components of the kinetic electron momentum $\hbar \bm{k} = - i \hbar \nabla + e \bm{A}$, where $e$ is the elementary positive charge, $\nabla$ is the Nabla operator, and $\bm{B} = \nabla \times \bm{A}$. For the vector potential, we choose a convenient gauge $\bm{A} = (B_y z - B_z y , - B_x z/2 , B_x y/2 )$, and we note that $k^2 = \bm{k} \cdot \bm{k}$.

The Hamiltonian of Eq.~(\ref{eq:ModelHamiltonianStart}) may be written in matrix form by projection onto a suitable set of basis states. In agreement with the boundary conditions, we use the basis states \cite{csontos:prb09}
\begin{equation}
\ket{j_z, n_z, n_y, \tilde{k}_x} =  \ket{j_z} \otimes \ket{\varphi_{n_z, n_y, \tilde{k}_x}} 
\label{eq:notationBasisStates}
\end{equation}
with orbital part
\begin{equation}
\varphi_{n_z, n_y, \tilde{k}_x}(x,y,z) = \frac{2}{\sqrt{L_z L_y}} \sin\Bigl[n_z \pi \left(\frac{z}{L_z} + \frac{1}{2} \right) \Bigr] \sin\Bigl[n_y \pi \left(\frac{y}{L_y} + \frac{1}{2} \right) \Bigr] e^{i \tilde{k}_x x} ,
\label{eq:functionOrbitalPart}
\end{equation} 
where the $n_z \geq 1$ and $n_y \geq 1$ are integer quantum numbers for the transverse subbands and $\tilde{k}_x$ is a wave number. Equation~(\ref{eq:functionOrbitalPart}) applies when both $|y| < L_y/2$ and $|z| < L_z/2$, otherwise $\varphi_{n_z, n_y, \tilde{k}_x} = 0$. The spin states $\ket{j_z}$ are eigenstates of $J_z$ and satisfy $J_z \ket{j_z} = j_z \ket{j_z}$, where $j_z \in \{3/2, 1/2, -1/2, -3/2\}$. In order to analyze the low-energy properties of $H$, we project it onto the 36-dimensional subspace with $n_z \leq 3$ and $n_y \leq 3$. This range of subbands is large enough to account for the most important couplings and small enough to enable fast numerical diagonalization. 

The band structure parameters of (bulk) Ge are \cite{lawaetz:prb71, winkler:book} $\gamma_1 = 13.35$, $\gamma_2 = 4.25$, $\gamma_3 = 5.69$, $\kappa = 3.41$, and $q = 0.07$, the deformation potential is \cite{birpikus:book} $b = -2.5\mbox{ eV}$. The values for the strain tensor elements $\epsilon_{xx} = -0.033 = \epsilon_{yy}$ and $\epsilon_{zz} = 0.020$ are obtained from finite element simulations, as described in the Supporting Information \cite{Supplement}. That is, the Ge lattice in the HW has almost completely adopted the lattice constant of Si along the $x$ and $y$ directions and experiences tensile strain along the out-of-plane direction $z$. Using moderate magnetic fields (of order Tesla) as in the experiment, $L_y = 20\mbox{ nm}$, $L_z \leq 3\mbox{ nm}$, and the above-mentioned parameters, we diagonalize the resulting 36$\times$36 matrix numerically and find that the eigenstates of lowest energy are close-to-ideal HH states. They feature spin expectation values $\langle J_z \rangle$ above 1.49 and below $-1.49$, respectively, when $\bm{B}$ is along $z$, and $\langle J_\nu \rangle \simeq 0$ for all $\nu \in \{x, y, z\}$ when $\bm{B}$ is in-plane. This corresponds to a LH admixture of less than 1\% \cite{Comment:LHAdmixture}. Furthermore, the admixture remains very small even when electric fields that may have been present in the experiment are added to the theory \cite{Supplement}.

The numerically observed HH character of the low-energy states in our model can easily be understood. First, with $\epsilon_{xx} = \epsilon_{yy} = \epsilon_\parallel$, the spin-dependent part of the strain-induced Hamiltonian can be written in the form $b(\epsilon_{zz} - \epsilon_\parallel) J_z^2$, and so basis states with $j_z = \pm 1/2$ are shifted up in energy by more than 250\mbox{ meV} compared to those with $j_z = \pm 3/2$. Second, the strong confinement along $z$ leads to an additional HH-LH splitting of the order of $\hbar^2 \pi^2 (m_{\rm LH}^{-1} - m_{\rm HH}^{-1})/(2 L_z^2)$, where $m_{\rm LH} = m/(\gamma_1 + 2 \gamma_2)$ and $m_{\rm HH} = m/(\gamma_1 - 2 \gamma_2)$. This results in a large splitting of $2 \gamma_2 \hbar^2 \pi^2 /(m L_z^2) \geq 710\mbox{ meV}$ for $L_z \leq 3\mbox{ nm}$. 

The result that hole states with $j_z = \pm 1/2$ are so much higher in energy than those with $j_z = \pm 3/2$ suggests that one may simplify the Hamiltonian of Eq.~(\ref{eq:ModelHamiltonianStart}) by projection onto the HH subspace, which is described in detail in the Supporting Information \cite{Supplement}. If the LH states are ignored, one expects small in-plane g-factors $g_\parallel \simeq 3 q \simeq 0.2$ and very large out-of-plane g-factors $g_\perp \simeq 6 \kappa + 27q/2 \simeq 21.4$ \cite{vankesteren:prb90, Supplement}. While $g_\parallel$ is indeed small in our experiment and $g_\perp \gg g_\parallel$ is indeed observed, the measured value of $g_\perp$ is significantly smaller than the one obtained from the pure-HH approximation. 

When we diagonalize the 36$\times$36 matrix, we find that the in-plane g-factors are close to $3q$, as also expected, e.g., from studies of the in-plane g-factors in narrow [001]-grown quantum wells \cite{vankesteren:prb90, winkler:prl00, winkler:book}. Our results for $g_\parallel$ agree well with the experiment and are consistent with the HH character of the low-energy states. Rather surprisingly, however, even though the low-energy eigenstates consist almost exclusively of either $\ket{3/2}$ or $\ket{-3/2}$ when the magnetic field is applied along $z$, we also find that the resulting $g_\perp \sim 15$ is indeed smaller than the value expected from the pure-HH approximation. The reason is that, in fact, the tiny admixtures from the LH bands are not negligible for the g-factors, as illustrated in Fig.~\ref{fig::ImageTheory} and described in the following. When the magnetic field is applied along the $z$ axis, the Zeeman-split states of lowest energy consist mostly of $\ket{-3/2, 1, 1, 0}$ and $\ket{3/2, 1, 1, 0}$, respectively. It turns out that the corresponding $g_\perp$ is strongly affected by the couplings  
\begin{equation}
C_{\pm} = \bra{\pm 3/2, 1, 1, 0} H \ket{\pm 1/2, 2, 2, 0} ,
\label{eq:dominantCoupling}
\end{equation}
because they satisfy $\left| C_{+} \right| \neq \left| C_{-} \right|$ in the presence of $B_z$ and therefore lead to different LH admixtures in the low-energy eigenstates of the HW \cite{Supplement}. The splitting between the basis states $\ket{\pm 1/2, 2, 2, 0}$ and $\ket{\pm 3/2, 1, 1, 0}$ in our model is predominantly determined by the confinement and can be approximated by $\Delta = \hbar^2 \pi^2 (4 m_{\rm LH}^{-1} - m_{\rm HH}^{-1})/(2 L_z^2)$ using $L_z \ll L_y$. From second-order perturbation theory \cite{winkler:book, Supplement}, we therefore find that the couplings of Eq.~(\ref{eq:dominantCoupling}) lead to a correction
\begin{equation}
g_C = \frac{\left|C_{-}\right|^2 - \left|C_{+}\right|^2}{\mu_B B_z \Delta} = 
- \frac{2^{17} \gamma_3^2}{81 \pi^4 (3 \gamma_1 + 10 \gamma_2)} 
\label{eq:gC}
\end{equation}
to the out-of-plane g-factor $g_\perp \simeq 6 \kappa + 27q/2 + g_C$. With the three Luttinger parameters $\gamma_{1,2,3}$ of Ge, this formula yields $g_C \simeq -6.5$, which is a substantial reduction of $g_\perp$ due to orbital effects \cite{csontos:prb09, vanbree:prb16}. 
Of course, $H$ couples $\ket{\pm 3/2, 1, 1, 0}$ not only with $\ket{\pm 1/2, 2, 2, 0}$ but also with other states. However, even when we take a large number of $10^4$ basis states into account ($n_y, n_z \leq 50$) and calculate the admixtures to $\ket{\pm 3/2, 1, 1, 0}$ via perturbation theory, we find that the sum of all corrections to $g_\perp$ is still close to $g_C$, i.e., Eqs.~(\ref{eq:dominantCoupling}) and (\ref{eq:gC}) describe the dominant part.

We note that if the HH-LH splitting in our model were dominated by the strain, such that $\Delta$ in Eq.~(\ref{eq:gC}) were much greater than the splitting caused by the confinement, the correction to $g_\perp$ from LH states would be suppressed and the model Hamiltonian would indeed approach the pure-HH approximation for the low-energy states \cite{Supplement}. Moreover, we found in our calculations that magnetic-field-dependent corrections to the g-factors are negligible given our HW parameters. This is consistent with $\sqrt{\hbar/(eB)} > L_y/2$ for $B \leq 6.5\mbox{ T}$, where $\sqrt{\hbar/(eB)}$ is the magnetic length, and agrees well with the experiment [see, e.g., Figure~\ref{fig::ImageII}~(o)]

While the result $g_\perp \sim 15$ from our simple model is already smaller than $g_\perp \sim 21$ from the pure-HH approximation, it is still larger than the measured values. We believe that this remaining deviation is mainly due to the following three reasons. First, given the small height of the HW, the eigenenergies in our model approach or even exceed the valence band offset \mbox{$\sim$0.5 eV} between Ge and Si \cite{Lu2005}, and so the hole wave function will leak into the surrounding Si. This certainly leads to a reduction of $g_\perp$, because the values of $\kappa$ in Ge and Si have opposite signs \cite{lawaetz:prb71, winkler:book}. Second, we used here the parameters of bulk Ge for simplicity. However, the strong confinement changes the gaps between the various bands of the semiconductor, which among other things may lead to a substantial rescaling of the effective band structure parameters \cite{winkler:book}. Improvements can be expected from an extended model that also involves the split-off band and the conduction band \cite{zielke:prb14, vanbree:prb16}. Finally, although our assumption of a long HW with a rectangular cross section is a reasonable approximation for the elongated HW QDs realized here, the details of the confinement along all three spatial directions can provide additional corrections. Taking all these elements fully into account is beyond the scope of the present work and requires extensive numerics.

In summary, having analyzed our HW model in detail, we can conclude that it reproduces all the key features of our experimental data and provides useful insight. It predicts a large g-factor anisotropy with $g_\parallel$ close to zero and $g_\parallel \ll g_\perp < 6\kappa$, as seen in the experiment. The spin projections calculated with our model suggest that the low-energy states of HWs are almost pure HHs and that the tiny admixtures from energetically higher LH states lead to a substantial reduction of $g_\perp$, which is a consequence of the orbital part of the magnetic-field-coupling. Finally, keeping in mind the finite potential barrier between Ge and Si, a possible explanation for the increasing $g_\parallel$ and the decreasing $g_\perp$ observed experimentally with increasing occupation number is that the confinement caused by the Ge/Si interface becomes less efficient as the eigenenergy of the hole increases (also due to the Coulomb repulsion which leads to an additional charging energy if more than one hole is present). Hence, a larger occupation number may change the effective aspect ratios of the HW QD experienced by the added hole and, thus, increase its HH-LH mixing.

The work was supported by the EC FP7 ICT project SiSPIN no. 323841, the EC FP7 ICT project PAMS no. 610446, the ERC Starting Grant no. 335497, the FWF-I-1190-N20 project and the Swiss NSF.
We acknowledge F.~Sch\"affler for fruitfull discussions related to the hut wire growth and for giving us access to the molecular beam epitaxy system, M. Schatzl for her support in electron beam lithography and V. Jadri\v{s}ko	for helping us with the COMSOL simulations.
Finally, we would like to thank G. Bauer for his continuous support.

\newpage

\section{Supporting Information}

\subsection{Finite element simulations of the strain in a HW}

\begin{figure}
\center
\includegraphics[scale = 1]{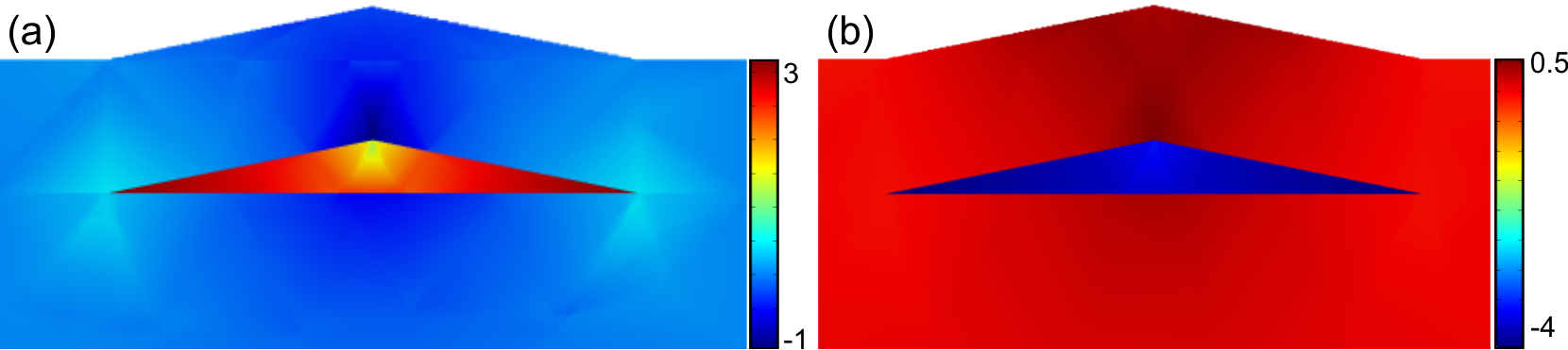}
	\caption{COMSOL simulations of the out-of-plane (a) and the in-plane strain distribution (b) in a capped HW. The color scale represents the percentage of strain with positive (negative) values meaning tensile (compressive) strain. } 
	\label{fig::SupImageI}
\end{figure}

The two images in Figure \ref{fig::SupImageI} represent COMSOL simulations of the out-of-plane (left) and the in-plane (right) strain distribution of a capped HW. For our theoretical model we have extracted an out-of-plane value of 2 and an in-plane value of -3.3 percent.

\subsection{Matrix representation of spin operators}

We use the following matrix representation \cite{winkler:book} for the operators $J_\nu$. The basis states are $\ket{3/2}$, $\ket{1/2}$, $\ket{-1/2}$, and $\ket{-3/2}$.
\begin{equation}
J_x = \begin{pmatrix}
0 & \frac{\sqrt{3}}{2} & 0 & 0 \\
\frac{\sqrt{3}}{2} & 0 & 1 & 0 \\
0 & 1 & 0 & \frac{\sqrt{3}}{2} \\
0 & 0 & \frac{\sqrt{3}}{2} & 0
\end{pmatrix}, \hspace{0.3cm}
J_y = \begin{pmatrix}
0 & - i \frac{\sqrt{3}}{2} & 0 & 0 \\
i \frac{\sqrt{3}}{2} & 0 & - i & 0 \\
0 & i & 0 & - i \frac{\sqrt{3}}{2} \\
0 & 0 & i \frac{\sqrt{3}}{2} & 0
\end{pmatrix}, \hspace{0.3cm}
J_z = \begin{pmatrix}
\frac{3}{2} & 0 & 0 & 0 \\
0 & \frac{1}{2} & 0 & 0 \\
0 & 0 & -\frac{1}{2} & 0 \\
0 & 0 & 0 & -\frac{3}{2}
\end{pmatrix}.
\label{eq:JmatricesSupplement}
\end{equation}
In the derivation of the pure-HH Hamiltonian [Eq.~(\ref{eq:ModelHamiltonianHH})], we consider the Pauli matrices
\begin{equation}
\sigma_x = \begin{pmatrix}
0 & 1 \\
1 & 0 \\
\end{pmatrix}, \hspace{1cm}
\sigma_y = \begin{pmatrix}
0 & -i \\
i & 0 \\
\end{pmatrix}, \hspace{1cm}
\sigma_z = \begin{pmatrix}
1 & 0 \\
0 & -1 \\
\end{pmatrix},
\label{eq:PauliMatrices}
\end{equation}
where $\ket{3/2}$ and $\ket{-3/2}$ are the basis states.

\subsection{Calculation with electric fields}

It is well possible that an electric field $E_z$ along the out-of-plane axis was present in the experiment. When the direct coupling $-e E_z z$ and the standard Rashba spin-orbit coupling $\alpha E_z (k_x J_y - k_y J_x)$, with $\alpha = -0.4\mbox{ nm$^2$} e$ \cite{winkler:book, Kloeffel2011}, are added to the Hamiltonian $H$ [Eq.~(\ref{eq:ModelHamiltonianStart}) of the main text], our finding that the low-energy states correspond to HH states remains unaffected, even for strong $E_z$ around 100\mbox{ V/$\mu$m}. Due to symmetries in our setup, we believe that electric fields $E_y$ along $y$ were very small. Nevertheless, we find numerically that the HH character of the eigenstates is preserved even when the direct and the standard Rashba coupling that are caused by nonzero $E_y$ are included in the model. We note that additional corrections besides the standard Rashba spin-orbit interaction arise for hole states in the presence of an electric field \cite{winkler:book}, but these terms are all small and will not change our result that the low-energy states are of HH type.

\subsection{Couplings $C_\pm$}

Here we explain the calculation of the matrix elements $C_\pm$ that are presented in Eq.~(\ref{eq:dominantCoupling}) of the main text. When the magnetic field is applied along the $z$ axis, the Hamiltonian is
\begin{eqnarray}
H &=& \frac{\hbar^2}{2m}\left[ 
\left(\gamma_1 + \frac{5 \gamma_2}{2}\right)k^2 
- 2 \gamma_2 \sum_\nu k_\nu^2 J_\nu^2
- 4 \gamma_3 \left( \{k_x, k_y\}\{J_x, J_y\} \mbox{ + c.p.} \right)
\right] \nonumber \\
& & + 2 \mu_B B_z \left( \kappa J_z + q J_z^3 \right) 
+ b \sum_\nu \epsilon_{\nu\nu} J_\nu^2 + V(y,z) 
\label{eq:ModelHamiltonianBzSupplement}
\end{eqnarray}  
and the vector potential is $\bm{A} = (- B_z y , 0 , 0 )$. Consequently,
\begin{eqnarray}
\{k_y , k_z \} &=& - \partial_y \partial_z , 
\label{eq:kykzSupplement} \\ 
\{k_x , k_z \} &=& - \partial_x \partial_z + i \frac{e}{\hbar} B_z y \partial_z , \label{eq:kxkzSupplement}\\
\{k_x , k_y \} &=& - \partial_x \partial_y + i \frac{e}{\hbar} B_z y \partial_y + i \frac{e}{2 \hbar} B_z , \\
k_x^2 &=& - \partial_x^2 + 2 i \frac{e}{\hbar} B_z y \partial_x + \frac{e^2}{\hbar^2} B_z^2 y^2 ,
\end{eqnarray}
and $k_y^2 = - \partial_y^2$, $k_z^2 = - \partial_z^2$. Using the matrices for the spin operators $J_\nu$ listed in Eq.~(\ref{eq:JmatricesSupplement}), one finds
\begin{eqnarray}
\bra{\pm 3/2} \{J_y, J_z \} \ket{\pm 1/2} &=& - i \frac{\sqrt{3}}{2}    , \\
\bra{\pm 3/2} \{J_x, J_z \} \ket{\pm 1/2} &=& \pm \frac{\sqrt{3}}{2}   ,
\end{eqnarray} 
whereas
\begin{equation}
\bra{\pm 3/2} Q \ket{\pm 1/2} = 0
\end{equation} 
when the operator $Q$ is $\{J_x, J_y \}$, $J_x^2$, $J_y^2$, $J_z^2$, $J_z$, or $J_z^3$. Therefore,
\begin{eqnarray}
C_{\pm} &=& \bra{\pm 3/2, 1, 1, 0} H \ket{\pm 1/2, 2, 2, 0}  \nonumber \\
&=& 
i \sqrt{3} \frac{\gamma_3 \hbar^2}{m}
\bra{\varphi_{1, 1, 0}} \{k_y, k_z\} \ket{\varphi_{2, 2, 0}}
\mp \sqrt{3} \frac{\gamma_3 \hbar^2}{m} 
\bra{\varphi_{1, 1, 0}} \{k_x, k_z\} \ket{\varphi_{2, 2, 0}} ,
\label{eq:CplusminusStartSupplement}
\end{eqnarray}
where the wave functions [see Eq.~(\ref{eq:functionOrbitalPart}) of the main text] of the basis states are
\begin{eqnarray}
\varphi_{1, 1, 0} &=& \frac{2}{\sqrt{L_z L_y}} \sin\Bigl[\pi \left(\frac{z}{L_z} + \frac{1}{2} \right) \Bigr] \sin\Bigl[\pi \left(\frac{y}{L_y} + \frac{1}{2} \right) \Bigr] , \\
\varphi_{2, 2, 0} &=& \frac{2}{\sqrt{L_z L_y}} \sin\Bigl[2 \pi \left(\frac{z}{L_z} + \frac{1}{2} \right) \Bigr] \sin\Bigl[2 \pi \left(\frac{y}{L_y} + \frac{1}{2} \right) \Bigr] 
\end{eqnarray} 
inside the HW ($|z| < L_z/2$, $|y| < L_y/2$) and $\varphi_{1, 1, 0} = 0 = \varphi_{2, 2, 0}$ outside. We note that $\bra{\varphi_{1, 1, \tilde{k}_x}} \partial_x \partial_z \ket{\varphi_{2, 2, \tilde{k}_x}}$ vanishes for arbitrary $\tilde{k}_x$ after integration over the $y$ axis due to the orthogonality of the basis functions for the $y$ direction. Thus, using Eqs.~(\ref{eq:kykzSupplement}) and (\ref{eq:kxkzSupplement}) in Eq.~(\ref{eq:CplusminusStartSupplement}) yields
\begin{equation}
C_{\pm} = 
-i \sqrt{3} \frac{\gamma_3 \hbar^2}{m}
\bra{\varphi_{1, 1, 0}} \partial_y \partial_z \ket{\varphi_{2, 2, 0}}
\mp i \sqrt{3} \frac{\gamma_3 e \hbar}{m} B_z 
\bra{\varphi_{1, 1, 0}} y \partial_z \ket{\varphi_{2, 2, 0}} .
\end{equation}
With the integrals (analogous for $z$)
\begin{eqnarray}
\int_{-L_y/2}^{L_y/2} dy \sin\Bigl[\pi \left(\frac{y}{L_y} + \frac{1}{2} \right) \Bigr] \frac{2 \pi }{L_y} \cos\Bigl[2 \pi \left(\frac{y}{L_y} + \frac{1}{2} \right) \Bigr] 
&=& - \frac{4}{3}, \\
\int_{-L_y/2}^{L_y/2} dy \sin\Bigl[\pi \left(\frac{y}{L_y} + \frac{1}{2} \right) \Bigr] y \sin\Bigl[2 \pi \left(\frac{y}{L_y} + \frac{1}{2} \right) \Bigr] 
&=& - \frac{8 L_y^2}{9 \pi^2},
\end{eqnarray}
we finally find
\begin{eqnarray}
\bra{\varphi_{1, 1, 0}} \partial_y \partial_z \ket{\varphi_{2, 2, 0}}
&=&  \frac{64}{9 L_y L_z} ,\\
\bra{\varphi_{1, 1, 0}} y \partial_z \ket{\varphi_{2, 2, 0}}
&=&  \frac{128 L_y}{27 \pi^2 L_z} ,
\end{eqnarray}
and so
\begin{equation}
C_{\pm} = 
-i \frac{64 \gamma_3 \hbar^2}{3 \sqrt{3}  L_y L_z m}
\mp i 
\frac{128 L_y \gamma_3 e \hbar B_z}{9 \sqrt{3} \pi^2 L_z m} .
\end{equation}
This is the result shown in Eq.~(\ref{eq:CpmResultSupplement}), considering that the Bohr magneton is $\mu_B = e \hbar / (2m)$. As explained in the above derivation, the first term on the right-hand side results from the part proportional to $\partial_y \partial_z \{ J_y, J_z \}$ in the Hamiltonian $H$, while the second term results from the part proportional to $B_z y \partial_z \{J_x, J_z \}$.

\subsection{Correction $g_C$ to the out-of-plane g-factor}

In the previous section we derived the couplings 
\begin{equation}
C_{\pm} = \bra{\pm 3/2, 1, 1, 0} H \ket{\pm 1/2, 2, 2, 0} =
-i \frac{64 \gamma_3 \hbar^2}{3 \sqrt{3}  L_y L_z m}
\mp i 
\frac{256 \gamma_3 L_y \mu_B B_z}{9 \sqrt{3} \pi^2 L_z}
\label{eq:CpmResultSupplement}
\end{equation}
assuming that the magnetic field is applied in the out-of-plane direction $z$.
In order to calculate the associated correction $g_C$ to the g-factor $g_\perp$, we consider a four-level system with the basis states $\ket{3/2, 1, 1, 0}$, $\ket{-3/2, 1, 1, 0}$, $\ket{1/2, 2, 2, 0}$, and $\ket{-1/2, 2, 2, 0}$ (see also Figure~\ref{fig::ImageTheory} (b) of the main article). Projection of the Hamiltonian $H$ [Eq.~(\ref{eq:ModelHamiltonianBzSupplement})] onto this basis yields the effective Hamiltonian
\begin{equation}
H_{\rm eff} = \begin{pmatrix}
E_{g,+} & 0 & C_{+} & 0 \\
0 & E_{g,-} & 0 & C_{-} \\
C_{+}^{*} & 0 & E_{e,+} & 0 \\
0 & C_{-}^{*}  & 0 & E_{e,-} 
\end{pmatrix} ,
\label{eq:HeffSupplement}
\end{equation}  
where the asterisk stands for complex conjugation and
\begin{eqnarray}
E_{g,\pm} &=& \frac{\hbar^2 \pi^2}{2 L_z^2 m_{\rm HH}} + \frac{\hbar^2 \pi^2 (\gamma_1 + \gamma_2)}{2 L_y^2 m} + \frac{9}{4} b (\epsilon_{zz} - \epsilon_\parallel) \nonumber \\
& & + \frac{(\pi^2 - 6) (\gamma_1 + \gamma_2) e^2 L_y^2 B_z^2 }{24 \pi^2 m} \pm \left( 3 \kappa + \frac{27}{4} q \right) \mu_B B_z   , 
\label{eq:EgpmSupplement} \\ 
E_{e,\pm} &=& \frac{2 \hbar^2 \pi^2}{L_z^2 m_{\rm LH}} + \frac{2 \hbar^2 \pi^2 (\gamma_1 - \gamma_2)}{L_y^2 m} + \frac{1}{4} b (\epsilon_{zz} - \epsilon_\parallel) \nonumber \\
& & + \frac{(2 \pi^2 - 3) (\gamma_1 - \gamma_2) e^2 L_y^2 B_z^2 }{48 \pi^2 m} \pm \left( \kappa + \frac{1}{4} q \right) \mu_B B_z 
\end{eqnarray}
are the energies on the diagonal. We assumed here that $\epsilon_{xx} = \epsilon_{yy} = \epsilon_\parallel$ and omitted the state-independent offset $15 b \epsilon_\parallel / 4$. The introduced effective masses are
\begin{eqnarray}
m_{\rm HH} &=& \frac{m}{\gamma_1 - 2 \gamma_2} , \\
m_{\rm LH} &=& \frac{m}{\gamma_1 + 2 \gamma_2} . 
\end{eqnarray}
From second-order perturbation theory \cite{winkler:book}, we find that the low-energy 2$\times$2 Hamiltonian obtained after diagonalization of Eq.~(\ref{eq:HeffSupplement}) is
\begin{equation}
H_{\rm eff}^{2\times 2} \simeq \begin{pmatrix}
E_{g,+} - \frac{\left| C_{+} \right|^2}{ \Delta_{+} } & 0 \\
0 & E_{g,-} - \frac{\left| C_{-} \right|^2}{ \Delta_{-} }
\end{pmatrix} ,
\label{eq:HeffgSupplement}
\end{equation} 
where we defined
\begin{equation}
\Delta_{\pm} = E_{e,\pm} - E_{g,\pm} .
\end{equation} 
With $\widetilde{\sigma}_z$ as a Pauli operator that is based on the low-energy eigenstates, Eq.~(\ref{eq:HeffgSupplement}) can be written as
\begin{equation}
H_{\rm eff}^{2\times 2} \simeq \frac{1}{2} \left( E_{g,+} +  E_{g,-} - \frac{\left| C_{+} \right|^2}{ \Delta_{+} } - \frac{\left| C_{-} \right|^2}{ \Delta_{-} } \right) + \frac{1}{2} \left( E_{g,+} - E_{g,-} - \frac{\left| C_{+} \right|^2}{ \Delta_{+} } + \frac{\left| C_{-} \right|^2}{ \Delta_{-} }  \right) \widetilde{\sigma}_z .
\end{equation}
The effective Zeeman splitting and the out-of-plane g-factor $g_\perp$ are therefore determined by
\begin{equation}
g_\perp \mu_B B_z \simeq  E_{g,+} - E_{g,-} - \frac{\left| C_{+} \right|^2}{ \Delta_{+} } + \frac{\left| C_{-} \right|^2}{ \Delta_{-} } .
\end{equation} 
From Eq.~(\ref{eq:EgpmSupplement}), it is evident that
\begin{equation}
E_{g,+} - E_{g,-} = \left( 6 \kappa + \frac{27}{2} q \right) \mu_B B_z .
\end{equation}
Given our parameters for Ge HWs, we find that the splittings $\Delta_\pm$ are predominantly determined by the confinement rather than the strain and that they can be well approximated by
\begin{equation}
\Delta_\pm \simeq  \frac{2 \hbar^2 \pi^2}{L_z^2 m_{\rm LH}} - \frac{\hbar^2 \pi^2}{2 L_z^2 m_{\rm HH}} = \frac{\hbar^2 \pi^2 (3 \gamma_1 + 10 \gamma_2)}{2 L_z^2 m} = \Delta 
\end{equation} 
using $L_z \ll L_y$. With the calculated expressions for the couplings $C_\pm$ [Eq.~(\ref{eq:CpmResultSupplement})], we finally obtain
\begin{equation}
g_\perp \simeq  6 \kappa + \frac{27}{2} q + g_C ,
\end{equation} 
where
\begin{equation}
g_C = \frac{\left| C_{-} \right|^2 - \left| C_{+} \right|^2}{\mu_B B_z \Delta } =  - \frac{2^{17} \gamma_3^2}{81 \pi^4 (3 \gamma_1 + 10 \gamma_2)} 
\end{equation} 
is the correction that results from the $B_z$-induced difference in the tiny LH admixtures ($\ket{\pm 1/2, 2, 2, 0}$) to the eigenstates of type $\ket{3/2, 1, 1, 0}$ and $\ket{-3/2, 1, 1, 0}$. We note that $|C_\pm|/\Delta < 0.05$ for our parameters, and so the perturbation theory used in the derivation of $H_{\rm eff}^{2\times 2}$ applies. Remarkably, our result for $g_C$ depends solely on the Luttinger parameters~$\gamma_{1,2,3}$.

\subsection{Hamiltonian for pure heavy holes}

If the contributions from LH states ($j_z = \pm 1/2$) are ignored completely, the Hamiltonian of Eq.~(\ref{eq:ModelHamiltonianStart}) in the main text can be simplified by projection onto the HH subspace, i.e., by removing all terms that cannot couple a spin $j_z = 3/2$ (or $j_z = -3/2$, respectively) with either $j_z = 3/2$ or $j_z = -3/2$. As evident, e.g., from the standard representations of the 4$\times$4 matrices $J_\nu$ and the 2$\times$2 Pauli matrices $\sigma_\nu$ [see Eqs.~(\ref{eq:JmatricesSupplement}) and (\ref{eq:PauliMatrices})], this projection can be achieved by substituting $\{J_x, J_y\} \to 0$ (analogous for cyclic permutations), $J_x^3 \to 3 \sigma_x /4$, $J_y^3 \to - 3 \sigma_y /4$, $J_z^3 \to 27 \sigma_z /8$, $J_{x,y}^2 \to 3/4$, $J_z^2 \to 9/4$, $J_{x,y} \to 0$, $J_z \to 3 \sigma_z /2$, which leads to the pure-HH Hamiltonian
\begin{eqnarray}
H_{\rm HH} &=& \frac{\hbar^2}{2m}\left[ 
\left(\gamma_1 - 2 \gamma_2 \right)k_z^2 
+ \left(\gamma_1 + \gamma_2\right)(k_x^2 + k_y^2)
\right] \nonumber \\
& & + \left( 3 \kappa + \frac{27}{4} q \right) \mu_B  B_z \sigma_z
+ \frac{3}{2} q \mu_B \left( B_x \sigma_x - B_y \sigma_y \right)
+ V(y,z) 
\label{eq:ModelHamiltonianHH}
\end{eqnarray}  
for the low-energy hole states in the HW. Thus, if LH states are ignored, one expects small in-plane g-factors $g_\parallel \simeq 3 q \simeq 0.2$ and very large out-of-plane g-factors $g_\perp \simeq 6 \kappa + 27q/2 \simeq 21.4$ \cite{vankesteren:prb90}.

\newpage

\newpage

\bibliography{bibliography}

\begin{mcitethebibliography}{45}
\bibitem[Com()]{Comment:Bir-Pikus}
In Eq.~(\ref{eq:ModelHamiltonianStart}) we omitted the term $-(a + 5 b
  /4)(\epsilon_{xx} + \epsilon_{yy} + \epsilon_{zz})$ from the Bir-Pikus
  Hamiltonian, where $a$ is the hydrostatic deformation potential. This term
  corresponds to a global energy shift in our model and therefore cannot affect
  the results. In an extended model, where the energy of the valence band edge
  matters, this term should be included.\relax
\mciteBstWouldAddEndPunctfalse
\mciteSetBstMidEndSepPunct{\mcitedefaultmidpunct}
{}{\mcitedefaultseppunct}\relax
\EndOfBibitem
\bibitem[Com()]{Comment:LHAdmixture}
An upper bound for the LH probability $p_{\rm LH}$ is given by $(1-p_{\rm LH})
  3/2 + p_{\rm LH} 1/2 > 1.49$, which results in $p_{\rm LH} < 0.01$.\relax
\mciteBstWouldAddEndPunctfalse
\mciteSetBstMidEndSepPunct{\mcitedefaultmidpunct}
{}{\mcitedefaultseppunct}\relax
\EndOfBibitem
\end{mcitethebibliography}

\end{document}